\documentclass[5p]{elsarticle}

\usepackage{color}

\usepackage{graphicx}
\usepackage{dcolumn}
\usepackage{bm}
\usepackage{hyperref}
\usepackage{amssymb}
\usepackage[normalem]{ulem}
\usepackage{multirow}

\begin{document}
\newcommand{\fig}[1]{Fig.~\ref{#1}}
\newcommand{\eq}[1]{eq.~(\ref{#1})}
\newcommand{\etal}{\textit{et al.}}

\title{Hydrogen-Water Mixtures in Giant Planet Interiors Studied with Ab Initio Simulations}
\author{F. Soubiran\corref{cor1}} 
\ead{fsoubiran@berkeley.edu}
\address{Department of Earth and Planetary Science, University of California, Berkeley, CA 94720, U.S.A.}

\author{B. Militzer} 
\address{Department of Earth and Planetary Science, Department of Astronomy, University of California, Berkeley, CA 94720, U.S.A.}

\date{\today}

\begin{abstract}

  We study water-hydrogen mixtures under
  planetary interior conditions using \textit{ab
    initio} molecular dynamics simulations. We determine the
  thermodynamic properties of various water-hydrogen mixing ratios
  at temperatures of 2000 and 6000~K for pressures of a few tens of
  GPa. These conditions are relevant for ice giant planets and for the
  outer envelope of the gas giants. We find that at 2000~K the mixture
  is in a molecular regime, while at 6000~K the dissociation of
  hydrogen and water is important and affects the thermodynamic
  properties.  We study the structure of the liquid and analyze the
  radial distribution function. We provide estimates for the transport
  properties, diffusion and viscosity, based on autocorrelation
  functions. We obtained viscosity estimates of the order of a few
  tenths of mPa$\,$s for the conditions under consideration. These
  results are relevant for dynamo simulations of ice giant planets.

\begin{description}
\item[Keywords:] hydrogen, water, planetary interior, transport properties.
\end{description}
\end{abstract}

\maketitle

\section{\label{sec:intro}Introduction}

The extraordinary discovery of more than one thousand confirmed
exoplanets in the past decade \cite{nasa_exoplanets} underlines the
necessity for a better description of their structure, formation, and
evolution. The discovery of an unexpectedly rich population of Neptune
and Sub-Neptune exoplanets \cite{petigura_2013} stressed the need for
a better understanding of these kinds of planets. Based on the
gravitational moments measured by Voyager II and orbital observations of their satellites \cite{jacobson_2007,jacobson_2009}, different models
have been proposed for the two ice giants of our solar system, Uranus
and Neptune \cite{helled_2011,nettelmann_2013}. Still, considerable
uncertainties about their internal composition remain, and a more
accurate equations of state (EOS) is needed to improve the models.

Furthermore, magnetic field observations provide additional information
and add some constraints on interior models. Dynamo simulations
\cite{stanley_2004,stanley_2006, soderlund_2013} predicted that water
in different phases is important for the magnetic field generation. 
So far, such dynamo simulations rely on ice giant interior models that
assume water to be fully phase separated from other compounds
including hydrogen. Recent {\it ab initio} computer simulations
predicted that water and metallic hydrogen readily mix at
pressure-temperature conditions in the cores of giant planets.
Experimental data obtained at lower pressures \cite{bali_2013}
predicted the true picture to be more complex and suggested partial
mixing of water and hydrogen. These results may also become relevant
for the interiors of gaseous giant planets like Saturn and Jupiter
because the presence of a small amount of water would affect the
physical properties of their envelopes. The properties of
hydrogen-water mixtures in different concentrations are thus of
significant interest in planetary science. 

Here we performed computer simulations of fluid water-hydrogen
mixtures at different concentrations under the pressure-temperature
conditions found in Neptune's and Uranus' upper mantle and the outer
envelope of Jupiter and Saturn. We explored the effects of the
concentration, temperature, and pressure on both the thermodynamic
properties and the structure of the fluid. The most significant
changes are introduced when the hydrogen and water dissociate at high
temperature and pressure. We also computed the transport properties,
such as particle diffusion and viscosity in order to better constrain
the dynamo simulations.

\section{\label{sec:methods}Simulation method}

We performed molecular dynamics simulations based on finite
temperature density functional theory (DFT) using the Vienna Ab initio
Simulation Package (VASP) \cite{vasp}. We studied four different
mixtures from H$_2$O:H$_2$=16:64, 24:48, 32:32, and 40:16 molecules in
simulation cell. We define the water concentration as
$x=N_{\textrm{H}_2\textrm{O}}/(N_{\textrm{H}_2\textrm{O}}+N_{\textrm{H}_2})$.

Simulation results for pure water and hydrogen can be found in
Refs.~\cite{french_2009,WilsonWongMilitzer2013,P21c} and
\cite{MC00,MC01,Vo07,Militzer2013,MH13}. Our simulations were at least
3~ps long for a time-step of 0.2~fs,
short enough to accurately describe the H-H and O-H bonds vibrations. The
temperature was kept constant using a Nos\'e thermostat
\cite{nose_1984, nose_1991}.

The electronic structure was computed at each time step using 
Mermin's finite temperature approach \cite{mermin_1965} of the
Kohn-Sham scheme \cite{kohn_1965}, with a Fermi-Dirac occupation
distribution. We used projector augmented wave (PAW) pseudopotentials
\cite{blochl_1994} with a cut-off radius of $r_\textrm{cut}=0.8~a_0$
for hydrogen and $r_\textrm{cut}=1.1~a_0$ for oxygen. For the
plane-wave basis, an energy cut-off of $1100$~eV was needed to reach a
convergence with less than 1\% inaccuracy on both the energy and the
pressure. We used the generalized gradient approximation (GGA) with
the Perdew, Burke, and Ernzerhof (PBE) \cite{perdew_1996}
exchange-correlation functional that gave reasonable results in
similar systems \cite{caillabet_2011,french_2009}. All simulations
were performed using periodic boundary conditions. We chose to use the
Baldereschi k-point \cite{baldereschi_1973} for sampling the Brillouin
zone because it gave similar thermodynamic functions (within 1\%) to a
2$\times$2$\times$2 Monkhorst-Pack grid \cite{monkhorst_1976}.

In this article we will present results at two temperatures: 2000 and
6000~K with pressures ranging from 10$-$35~GPa and 10$-$70~GPa for the
lower and higher temperatures, respectively. These conditions are
close to what we find in Neptune's or Uranus' upper mantle
\cite{hubbard_1991,podolak_1995,nettelmann_2013} 
or in the outer envelop of Jupiter and Saturn
\cite{leconte_2012,MHVTB}.

\section{\label{sec:res}Results and discussion}

\subsection{\label{sec:thermo} Thermodynamic properties}

\begin{figure}[ht]
 \centering
 \includegraphics[width=\columnwidth]{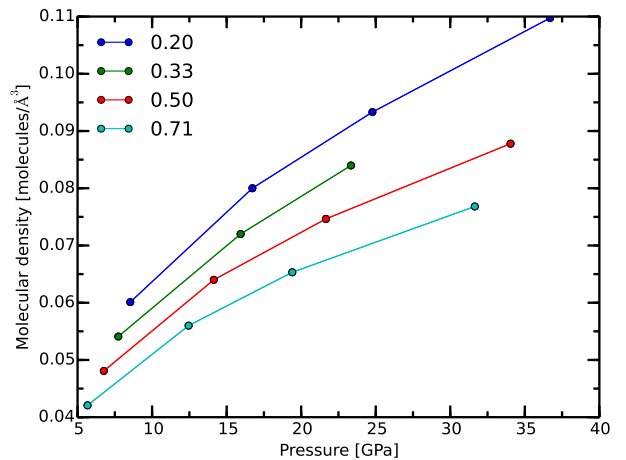}
 \caption{\label{fig:moldensvsP_2kK} (color online) Molecular density as a function of the pressure at 2000~K. The four colors are for the four different
   H$_2$O concentrations, $x$, given in the caption. }
\end{figure}

We extracted the pressure and the internal energy time-averages from
the different molecular dynamics simulations. The error bars presented
in the figures were computed using the block
averaging method \cite{rapaportbook}. In \fig{fig:moldensvsP_2kK}, we
report the evolution of the molecular density as a function of the
pressure at 2000~K. The trend is quite similar for
each mixing ratio.  Molecular densities are much higher for
hydrogen-rich mixtures, which simply reflects the fact the hydrogen
molecules are smaller than the water molecules.

\begin{figure}[ht]
 \centering
 \includegraphics[width=\columnwidth]{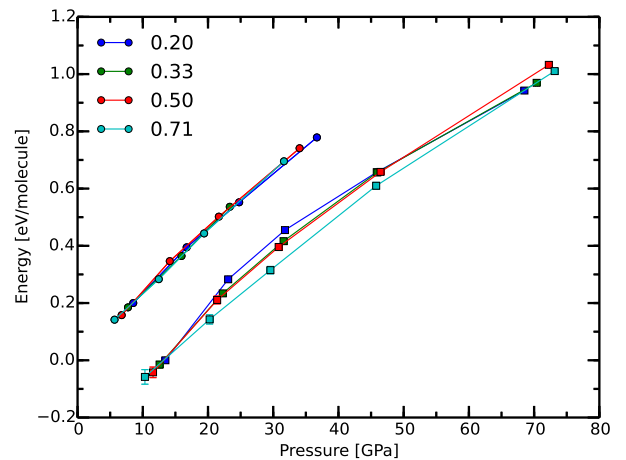}
 \caption{\label{fig:EvsP_2+6kK} (color online) Internal energy per
   molecule as a function of the pressure at 2000~K ($\bullet$) and
   6000~K ($\blacksquare$). The energies have been manually shifted for each
   concentration and temperature to better compare them as function of
   pressure. At 2000~K, the energy origins for each concentration have been chosen so that at 8.6 GPa, the
   energy is 0.2~eV/mol.  For the 6000 K isotherm, the energy at
   13.5~GPa is 0.0~eV/mol. The four colors are for the four different
   H$_2$O concentrations, $x$, given in the caption. }
\end{figure}

In \fig{fig:EvsP_2+6kK}, we plotted the internal energy per molecule
as function of pressure for two temperatures. The curves for each
concentration have been shifted in order to superpose them for each
isotherm and compare their evolution as a function of the pressure.

At 2000~K, the linear behavior of the energy is very similar for all
concentrations, indicating that the thermal excitations of H$_2$ and
H$_2$O molecules make similar contributions to both the internal
energy and the pressure. In contrast, we find significant deviations
from the linear relationship at 6000~K. More importantly, there is a
dependence on the concentration, indicating a change of structure of
the liquid, which we discuss in section~\ref{sec:struct}.

\subsection{\label{sec:struct} Structure and composition}

\begin{figure}[ht]
 \centering
 \includegraphics[width=\columnwidth]{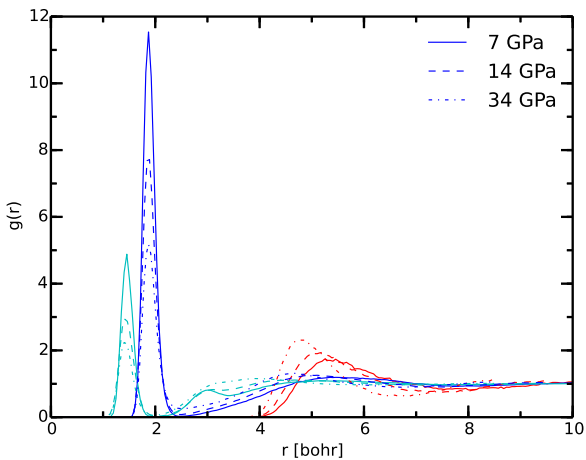}
 \caption{\label{fig:32-32_RDF_2000K} (color online) Radial distribution function of a $x=0.5$ mixture at 2000~K and different pressures. The O-O structure is plotted in red, O-H in blue and H-H in cyan. }
\end{figure}

The molecular dynamics simulations provide information about the
microscopic structure of the fluid. Radial distribution functions
(RDF) are an established tool \cite{frenkelbook,hansenbook} to analyze
the structure. A first example is shown in \fig{fig:32-32_RDF_2000K},
where we plotted the RDF of an equimolar mixture, $x=0.5$, at three
different pressures and a 2000~K temperature. The
large peaks in the O-H and H-H correlations underline
the molecular character of the fluid, which is fully molecular up to
at least 35~GPa.  There appear to be no other strong correlations in
the fluid besides the molecular structure except at highest pressure
where the correlation among the oxygen nuclei increases. Under these
conditions, the fluid is close to
phase transition to a superionic regime at 40-50~GPa at 2000 K in pure
water \cite{french_2009}.

\begin{figure}[ht]
 \centering
 \includegraphics[width=\columnwidth]{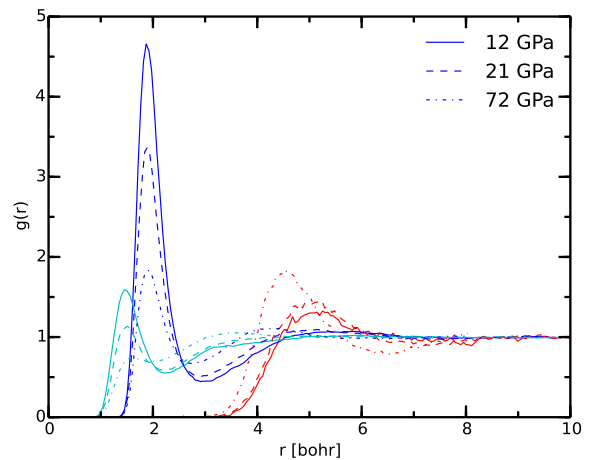}
 \caption{\label{fig:32-32_RDF_6000K} (color online) Radial distribution function of a $x=0.5$ mixture at 6000~K and different pressures. The O-O structure is plotted in red, O-H in blue and H-H in cyan. }
\end{figure}

In \fig{fig:32-32_RDF_6000K}, we plotted the RDF for a $x=0.5$ mixture
at 6000~K. The size of the H-H and H-O peaks is significantly reduced
compared to 2000~K indicating a partial dissociation of the molecules.
This dissociation is consistent with the behavior observed in pure
hydrogen \cite{caillabet_2011} and water \cite{french_2009}. We infer
that the dissociation is the cause of the concentration dependence of
the internal energy-pressure relationship in \fig{fig:EvsP_2+6kK}.
We did not identify any other correlations besides the molecular peaks but 
it should be noted that the O-O correlation increases at 70 GPa.

\subsection{\label{sec:diffusion} Transport properties}
\subsubsection{Autocorrelation functions}

We derived the viscosity and diffusion coefficients of the hydrogen
and oxygen species in each mixture in order to provide some dynamic
information about the mixture and ultimately to improve models for
planetary interiors. We used the autocorrelation functions for the
diffusion and viscosity calculations. The diffusion coefficient of
species is given by the Green-Kubo formula
\cite{danel_2012,frenkelbook}:

\begin{equation}
 D_\alpha=\frac{1}{3N_\alpha}\sum_\alpha\int_0^{+\infty}\langle\textbf{v}_\alpha(\tau)\cdot\textbf{v}_\alpha(0)\rangle \;\textrm{d}\tau,
\end{equation}
where we sum over all $N_\alpha$ nuclei of type $\alpha$.
$\textbf{v}_\alpha(\tau)$ is the particle's velocity vector at time
$\tau$. The brackets mean that we average over many possible time
origins assuming the fluid has reached an equilibrium in our
simulations\cite{hailebook}. To reduce the fluctuations further, we
averaged the result from different maximum integration times ranging
from 75 to 100\% of our time-window. The error bars in Figures 
\ref{fig:D_OvsP_2+6kK}-\ref{fig:D_HvsP_2+6kK} and Tables 
\ref{table:2kK}-\ref{table:6kK} represent one standard deviation from the mean.

We computed the viscosity, $\eta$, from the autocorrelation function of the stress-tensor $\sigma_{\alpha\beta}$,
\begin{equation}
 \eta=\frac{V}{3k_\textrm{B}T}\sum_{\{\alpha\beta\}}\int_0^{+\infty}\langle\sigma_{\alpha\beta}(\tau)\sigma_{\alpha\beta}(0)\rangle \;\textrm{d}\tau,
\end{equation}
where the sum $\{\alpha\beta\}$ includes all the deviatoric stress
tensor components, $\{xy,xz,yz\}$. $V$ is the volume of the
simulation cell and $T$ the temperature of the system. There are still
strong fluctuations in the stress tensor autocorrelation function even
when we use multiple origins because the stress-tensor is a bulk
property and it is not possible to average over the multiple particles.  In
order to reduce the spreads in the results we fitted the
autocorrelation function by a simple exponential function. This method
provides more stable estimates of the viscosity. Nevertheless, we do
not provide error bars in the figures because our viscosity values
have to be considered to be order-of-magnitude estimates only because
it is too difficult to obtain an accurate determination of the viscosity
and its error bar.

\subsubsection{Diffusion}

We analyze the diffusion of individual nuclei rather than tracking the
motion of molecules because the latter method becomes difficult as
soon as dissociation and recombination effects occur at elevated
temperature.

\begin{figure}[ht]
 \centering
 \includegraphics[width=\columnwidth]{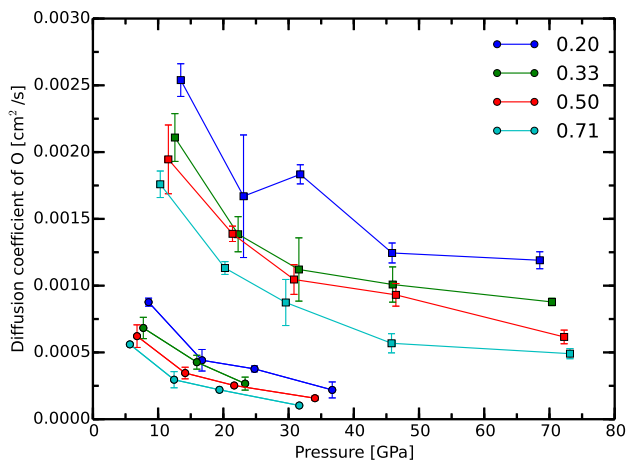}
 \caption{\label{fig:D_OvsP_2+6kK} (color online) Oxygen diffusion coefficient as a function of the pressure at 2000~K ($\bullet$) and 6000~K ($\blacksquare$).The four colors are for the four different
   H$_2$O concentrations, $x$, given in the caption.  }
\end{figure}

In \fig{fig:D_OvsP_2+6kK}, we plotted the diffusion coefficient of
oxygen nuclei as function of pressure for 2000 and 6000~K. For both
temperatures, we find that diffusion slows down with increasing
pressure because the particle interact more strongly. At 6000~K, the
diffusion is faster because the additional kinetic energy allows
particles to hop over potential barriers more rapidly. The oxygen
diffusion rate decreases with increasing water concentration
presumably because water molecules interact more strongly with each
other than they do with hydrogen molecules at the same pressure.

\begin{figure}[ht]
 \centering
 \includegraphics[width=\columnwidth]{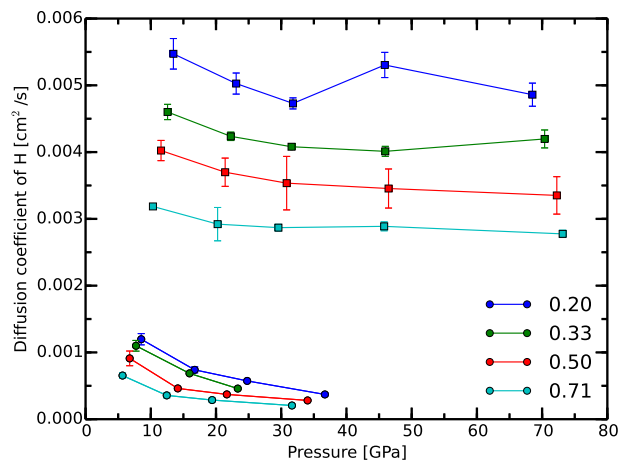}
 \caption{\label{fig:D_HvsP_2+6kK} (color online) Hydrogen diffusion coefficient as a function of the pressure at 2000~K ($\bullet$) and 6000~K ($\blacksquare$). The four colors are for the four different
   H$_2$O concentrations, $x$, given in the caption.  }
\end{figure}

The diffusion coefficient of the hydrogen nuclei are plotted in
\fig{fig:D_HvsP_2+6kK}. For the dissociated fluids at 6000~K, the
hydrogen atoms diffuse approximately twice as fast as the oxygen atoms
that are heavier and larger. In the molecular regime, the hydrogen and
oxygen diffusion rates are more similar because some hydrogen nuclei
are bound to oxygen. Again we find that when the water concentration
rises, the diffusion slows down. There is at least two reasons for
that.  First, an increasing fraction of hydrogen nuclei is bound in
water molecules, which greatly restricts their mobility. Second the
mobility of the hydrogen molecules is reduced by the interaction with
water molecules. 

It is interesting to note that the diffusion rates at 6000~K are
almost independent of pressure. This is most likely due to two
counter-acting effects. With increase pressure, particles interact
more strongly, which reduces diffusion rates. A pressure increase also
introduces a rise in the dissociation fraction, which increases the
mobility of different nuclei.

\subsubsection{Viscosity}
\begin{figure}[ht]
 \centering
 \includegraphics[width=\columnwidth]{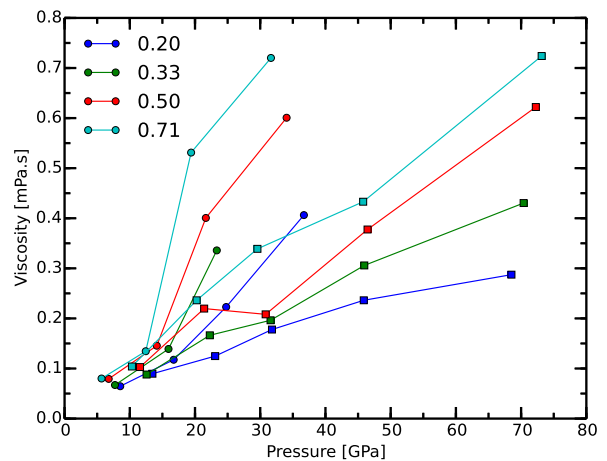}
 \caption{\label{fig:viscosityvsP_2+6kK} (color online) Viscosity as a function of the pressure at 2000~K ($\bullet$) and 6000~K ($\blacksquare$). The four colors are for the four different
   H$_2$O concentrations, $x$, given in the caption.  }
\end{figure}

In \fig{fig:viscosityvsP_2+6kK}, we plotted our viscosity estimates
for the H$_2$-H$_2$O mixtures. Values vary from 0.1 to 1~mPa$\,$s over
the range of parameters that we explored here. The viscosity rises
with increasing water concentration because water molecules interact
more strongly than hydrogen molecules at the same pressure. The
viscosity rises with increasing pressure and decreasing temperature, as
expected. Viscosity estimates are of importance for the
numerical simulations of planetary interiors, especially for
Neptune-like planets where the water content is presumably high.
Moreover, the viscosity plays a prominent role in the dynamics of the
dynamo \cite{stanley_2004,stanley_2006, soderlund_2013} and reliable
estimates are therefore needed.

\begin{table}[h]
\caption[Table caption text]{Hydrogen and oxygen diffusion coefficients and pressure table for different molecular densities and water concentrations $x_{{\textrm{H}_2\textrm{O}}}$ at 2000~K.}
\label{table:2kK}
{\scriptsize
\begin{tabular}{ccccc}
\hline
\multirow{2}{*}{$x_{{\textrm{H}_2\textrm{O}}}$} & Mol. dens. & Pressure & H diffusion & O diffusion \\
 & ($\times 10^{-2} $\AA$\!^{-3}$) &(GPa) & ($\times 10^{-3}\!$~cm$^2$/s) & ($\times 10^{-3}\!$~cm$^2$/s) \\
\hline
0.20 & 6.0105 & 8.534(36)& 1.197(84)& 0.876(30)\\
0.20 & 8.0000 & 16.73(11)& 0.739(45)& 0.441(80)\\
0.20 & 9.3308 & 24.776(45)& 0.573(10) & 0.377(22)\\
0.20 & 10.974 & 36.694(75)& 0.371(23)& 0.219(59)\\
0.33 & 5.4095 & 7.731(46)& 1.099(80)& 0.683(90)\\
0.33 & 7.2000 & 15.936(56)& 0.686(14)& 0.427(52)\\
0.33 & 8.3977 & 23.339(90)& 0.459(16)& 0.266(49)\\
0.50 & 4.8084 & 6.763(30)& 0.91(11)& 0.622(85)\\
0.50 & 6.4000 & 14.139(64)& 0.461(19)& 0.346(43)\\
0.50 & 7.4646 & 21.66(13)& 0.371(15)& 0.252(7)\\
0.50 & 8.7792 & 34.04(14)& 0.281(12)& 0.157(17)\\
0.71 & 4.2074 & 5.667(51)& 0.654(39)& 0.560(14)\\
0.71 & 5.6000 & 12.45(11)& 0.357(38)& 0.296(61)\\
0.71 & 6.5316 & 19.40(14)& 0.287(32)& 0.060(18)\\
0.71 & 7.6818 & 31.64(10)& 0.205(23)& 0.090(13)\\
\hline
\end{tabular}}
\end{table}

\begin{table}[h]
\caption[Table caption text]{Same as Table \ref{table:2kK} but at 6000~K. }
\label{table:6kK}
{\scriptsize
\begin{tabular}{ccccc}
\hline
\multirow{2}{*}{$x_{{\textrm{H}_2\textrm{O}}}$}& Mol. dens. & Pressure & H diffusion & O diffusion \\
 & ($\times 10^{-2} $\AA$\!^{-3}$) &(GPa) & ($\times 10^{-3}\!$~cm$^2$/s) & ($\times 10^{-3}\!$~cm$^2$/s) \\
\hline
0.20 & 6.0105 & 13.47(11)& 5.47(23)& 2.54(12)\\
0.20 & 8.0000 & 23.09(11)& 5.03(16)& 1.67(46)\\
0.20 & 9.3308 & 31.80(12)& 4.729(83)& 1.834(72)\\
0.20 & 10.974 & 45.87(11)& 5.30(19)& 1.245(75)\\
0.20 & 13.027 & 68.514(96)& 4.86(17)& 1.190(64)\\
0.33 & 5.4095 & 12.57(12)& 4.60(11)& 2.11(18)\\
0.33 & 7.2000 & 22.26(11)& 4.236(63)& 1.39(13)\\
0.33 & 8.3977 & 31.581(93)& 4.080(34)& 1.12(24)\\
0.33 & 9.8765 & 45.95(16)& 4.011(75)& 1.01(13)\\
0.33 & 11.724 & 70.38(16)& 4.20(13)& 0.878(64)\\
0.50 & 4.8084 & 11.570(81)& 4.02(15)& 1.95(26)\\
0.50 & 6.4000 & 21.40(18)& 3.70(21)& 1.388(56)\\
0.50 & 7.4646 & 30.83(15)& 3.54(40)& 1.05(11)\\
0.50 & 8.7792 & 46.45(16)& 3.46(29)& 0.931(85)\\
0.50 & 10.421 & 72.25(16)& 3.35(28)& 0.616(51)\\
0.71 & 4.2074 & 10.32(11)& 3.186(32)& 1.76(10)\\
0.71 & 5.6000 & 20.24(17)& 2.92(25)& 1.132(48)\\
0.71 & 6.5316 & 29.57(16)& 2.869(50)& 0.87(17)\\
0.71 & 7.6818 & 45.78(19)& 2.888(67)& 0.568(72)\\
0.71 & 9.1187 & 73.14(23)& 2.775(54)& 0.491(38)\\
\hline
\end{tabular}}
\end{table}

\section{\label{sec:conclusion}Conclusion}

We performed {\it ab initio} computer simulations of H$_2$-H$_2$O
mixtures at 2000~K and 6000~K in order to provide information about
the physics of such systems under conditions in the interior of icy or
gaseous giant planets. We showed for instance that at 2000~K and a few
tens of GPa, the mixture is purely molecular and its thermodynamic
behavior depends only weakly on the water concentration. In contrast,
at 6000~K we observe partial dissociation of both H$_2$ and H$_2$O
molecules, which changes in the thermodynamic behavior.  Because
dissociation is often associated with a reduction in the electronic
band gap, we can infer that the dissociated fluids~\cite{collins_2001}
have a significantly higher electronic conductivity. This has
potential implications for the magnetic field generation in water-rich
gas giants planets.

The autocorrelation function analysis provided robust results for the
diffusion of oxygen and hydrogen nuclei in the mixtures.  The
diffusion of hydrogen is almost constant at 6000~K from 10 to 70~GPa
range due to the dissociation process that add to the mobility of the
protons. The diffusion coefficients are important for the planetary
models because the diffusion influences the convective behavior in the
semi-convective regime as shown by Leconte and Chabrier
\cite{leconte_2012}. We also computed the viscosity which is of the
order of a few tenths of mPa$\,$s at the conditions under
consideration. This is not a very viscous fluid indicating a turbulent
behavior in planetary interiors.

\section{Acknoledgements}
This work has been supported by NSF and NASA. The simulations were
performed on the NASA supercomputing facilities.

\bibliographystyle{elsarticle-num-names.bst}
\section*{References}
\bibliography{HEDLA}

\end{document}